\documentclass[twocolumn,showpacs,preprintnumbers,amsmath,amssymb]{revtex4-1}

\usepackage{graphicx}
\usepackage{dcolumn}
\usepackage{bm}
\usepackage[dvipsnames]{xcolor}
\usepackage{color}
\usepackage{mathrsfs}
\usepackage{rotating}
\usepackage{amsfonts}
\usepackage[american]{babel}
\usepackage{pstricks}
\usepackage{pstricks-add}
\usepackage{pstricks}
\usepackage{pst-plot}
\usepackage{pst-grad}
\usepackage{pst-eps}
\usepackage{amssymb}

\newcommand{\bea}{\begin{eqnarray}}
\newcommand{\eea}{\end{eqnarray}}
\newcommand{\beq}{\begin{equation}}
\newcommand{\eeq}{\end{equation}}

\begin{document}

\title{Quasinormal modes of regular black holes}

\author{Antonino Flachi and Jos\'e P. S. Lemos}
\affiliation{Centro Multidisciplinar de Astrof\'{\i}sica - CENTRA,
Departamento de F\'{\i}sica, Instituto Superior T\'ecnico - IST,
Universidade T\'{e}cnica de Lisboa - UTL,\\ Av. Rovisco Pais 1, 1049-001
Lisboa, Portugal\\
Emails: antonino.flachi@ist.utl.pt, joselemos@ist.utl.pt}

\date{\today}

\begin{abstract}
Black hole quasinormal frequencies are complex numbers that encode
information on how a black hole relaxes after it has been perturbed
and depend on the features of the geometry and on the type of
perturbations. On the one hand, the examples studied so far in the
literature focused on the case of black hole geometries
with singularities in their interior. On
the other hand, it is expected that quantum or classical modifications
of general relativity may correct the pathological singular behavior
of classical black hole solutions. Despite the fact that we do not
have at hand a complete theory of quantum gravity, regular black hole
solutions can be constructed by coupling gravity to an external form
of matter, sometimes modeled by
one form or another of nonlinear electrodynamics. It
is therefore relevant to compute quasinormal frequencies for these
regular solutions and see how differently, from the ordinary ones,
regular black holes ring. In this paper, we take a step in this direction and, by computing the quasinormal frequencies, study the quasinormal modes of neutral and charged scalar field perturbations on regular black hole backgrounds in a variety of models.
\end{abstract}


\maketitle


\section{Introduction}
\label{sec1}

The problem of understanding how avoidance of singularities may be
possible in general relativity goes beyond formal importance and is
not at all new. In 1968 Bardeen constructed the first example of a
regular black hole,  i.e., a regular, nonsingular geometry
with an event horizon, satisfying the weak energy conditions
\cite{bardeen}.
Despite its theoretical relevance, Bardeen's
regular black hole solution
lacked, for several years, a satisfactory physical
interpretation. The reason is that it is not a vacuum solution of
Einstein's equations and, in order to generate it, it is necessary to
introduce some external form of matter or a modification to
gravity. The solution was obtained by introducing an
{\it ad hoc} energy momentum tensor, regular and bounded, decaying at
infinity, and satisfying the weak energy conditions. However, no
fundamental physical motivation for this choice had been given,
until Ayon-Beato and Garcia reobtained
the solution by describing it as the gravitational
field of some sort of nonlinear
magnetic monopole \cite{AyonBeato:2000zs}.

Bardeen's work has been followed by several
other examples, motivating deeper analyses of how singularity
avoidance may be possible in general.
Other solutions have been proposed in the literature and some
attention has been directed to theories of gravity coupled to
nonlinear electrodynamics.
Solutions have been discussed in different contexts, and a few
of interest to us are, in addition to
Refs.~\cite{bardeen,AyonBeato:2000zs}, those analyzed in
Refs.~\cite{sh,bronnikov,berej,dymn2,fabris,ayonbeato}.
It is worth mentioning that an important viable example of a
black hole with a regular center was constructed by Dymnikova,
with a de Sitter core smoothly connecting to a
Schwarzschild outer geometry \cite{dymn}.
Analysis of regular black hole solutions continued in several
directions, see, e.g., Refs.~\cite{Ansoldi:2008jw,Lemos:2011dq}.
Another
important step toward understanding the absence of singularities
in general relativity was taken by Borde who showed that, for a large
class of black hole solutions, absence of singularities was related to
a change in the topology beyond the event horizon
\cite{Borde:1996df}. Borde's arguments clearly demonstrated the
impossibility of proving general singularity theorems when the strong
energy condition or the existence of global hyperbolicity were not
assumed.

Now, the interior is a hidden region by definition.
The region that connects the interior to the exterior is
the horizon and its near horizon region.
Thus, a way to peer into the interior is to
perturb the horizon.
More important, quantum processes
occur in its neighborhood, giving a glimpse of phenomena that
unites quantum mechanics and gravitation. Indeed,
the Hawking radiation has its origin in the vicinity
of the horizon and the black hole entropy
is believed to come out from degrees of freedom located
at the horizon. Thus, the horizon is the region that
may give a glimpse of what the black hole interior is.
By poking the horizon something from the inside
might pop up.

One way to poke the horizon is to perturb spacetime, create
quasinormal modes (QNMs), and analyze the result
of the perturbation.  Quasinormal
frequencies (QNFs) are complex numbers that encode information on the
system's relevant parameters and on its relaxation after it has been
perturbed. Quasinormal modes are related to the classical evolution of
a system.  They can also reveal instability.
A way to study QNMs is through a WKB approximation as was
done first in \cite{SchutzWill}. This was further developed by Iyer
and Will \cite{iw,iyer2} to study the Schwarzschild black hole and by
\cite{leaver,Hod:1997mt,Konoplya:2002ky,kokkotasberti,neitz,Konoplya:2011qq}
to study perturbation in the Reissner-Nordstr\"om case as well as
charged scalar perturbations. Modes with
large imaginary parts were analyzed, in connection
with a possible horizon area quantization law, in
Schwarzschild and Reissner-Nordstr\"om see
\cite{Motl:2003cd,Motl:2002hd}.  For reviews see
\cite{Nollert:1999ji,Kokkotas:1999bd,Berti:2009kk}.

All this work has been done for vacuum black holes.
It is important to proceed for regular black holes.
A first step was made by Fernando and Correa \cite{fernando}
who computed the QNFs for Bardeen's solution
\cite{bardeen,AyonBeato:2000zs}.
In addition in Ref.~\cite{bronkonzid} a study of the QNMs for the
 solution presented in [8] was performed.
In this paper, we intend to study QNFs for the solutions given in
Refs.~\cite{bardeen,AyonBeato:2000zs,sh,bronnikov,berej,dymn2,fabris,ayonbeato}.

\vspace{-0.3cm}
\section{Some regular black holes}
\label{bardeen}

In the above context of gravity coupled to some form of matter,
regular solutions were obtained from a prototypical action of the form
\bea
S= {1\over 16\pi}\int d^4x\sqrt{-g}\left( R -\mathscr{L}\right)~,
\eea
where $g$ is the determinant of the metric $g_{\mu\nu}$,
$R$ is the scalar curvature, and $\mathscr{L}$
represents the Lagrangian of the matter fields.
For the case of nonlinear electrodynamics then
$\mathscr{L} = \mathscr{L}\left(F\right)$ is a
nonlinear function of the electromagnetic field strength with
$F=F_{\mu\nu}F^{\mu\nu}/4$.

The general line element
\bea
ds^2 = g_{\mu\nu}dx^\mu dx^\nu\,,
\label{generalmet}
\eea
when presented for spherically symmetric
regular black hole solutions takes the
form
\bea
ds^2 = - f(r) dt^2 + \frac{dr^2}{f(r)} +r^2
\left(d\theta^2+\sin^2\theta\,d\phi^2\right)~,
\label{rbhm}
\eea
where $(t,r,\theta,\phi)$
are the usual space-time spherical coordinates and
the lapse function $f\equiv f(r)$ depends on the specific form
of underlying matter.

In \cite{bardeen}
the function takes a particularly simple form
\bea
f(r) = 1-\frac{2mr^2}{(r^2+\alpha^2)^{3/2}}\,,
\label{bard}
\eea
with $\alpha=
{\rm const}$ and $m$ being the mass
of the solution.
This implies a specific matter energy-momentum tensor
that is de Sitter at the core and vanishes away at infinity
as a magnetically charged solution in the context of a specific
nonlinear electrodynamics with Lagrangian given by
$\mathscr{L}=(3/(2s\alpha^2))(\sqrt{2\alpha^2F}/(1+\sqrt{2\alpha^2F}))^{5/2}$,
where $s=|\alpha|/2m$ and $\alpha$ is the magnetic charge
\cite{AyonBeato:2000zs}. Depending on the relative values of $m$ and
$\alpha$, Eq.~(\ref{hay}) can have two, one or zero horizons.

In \cite{sh} the function also takes a simple form
\bea
f(r) = 1-\frac{2mr^2}{r^3+2\alpha^2}\,
\label{hay}
\eea
which is a variant from the original Bardeen's proposal
\cite{bardeen} with $\alpha=
{\rm constant}$.
This also implies a specific matter energy-momentum tensor
that is de Sitter at the core and vanishes away at infinity.
As well, depending on the relative
values of $m$ and $\alpha$ equation (\ref{hay})
can have two, one or zero horizons \cite{sh}.

\begin{figure}[ht]
\unitlength=1.1mm
\begin{picture}(110,40)
\includegraphics[scale=0.35]{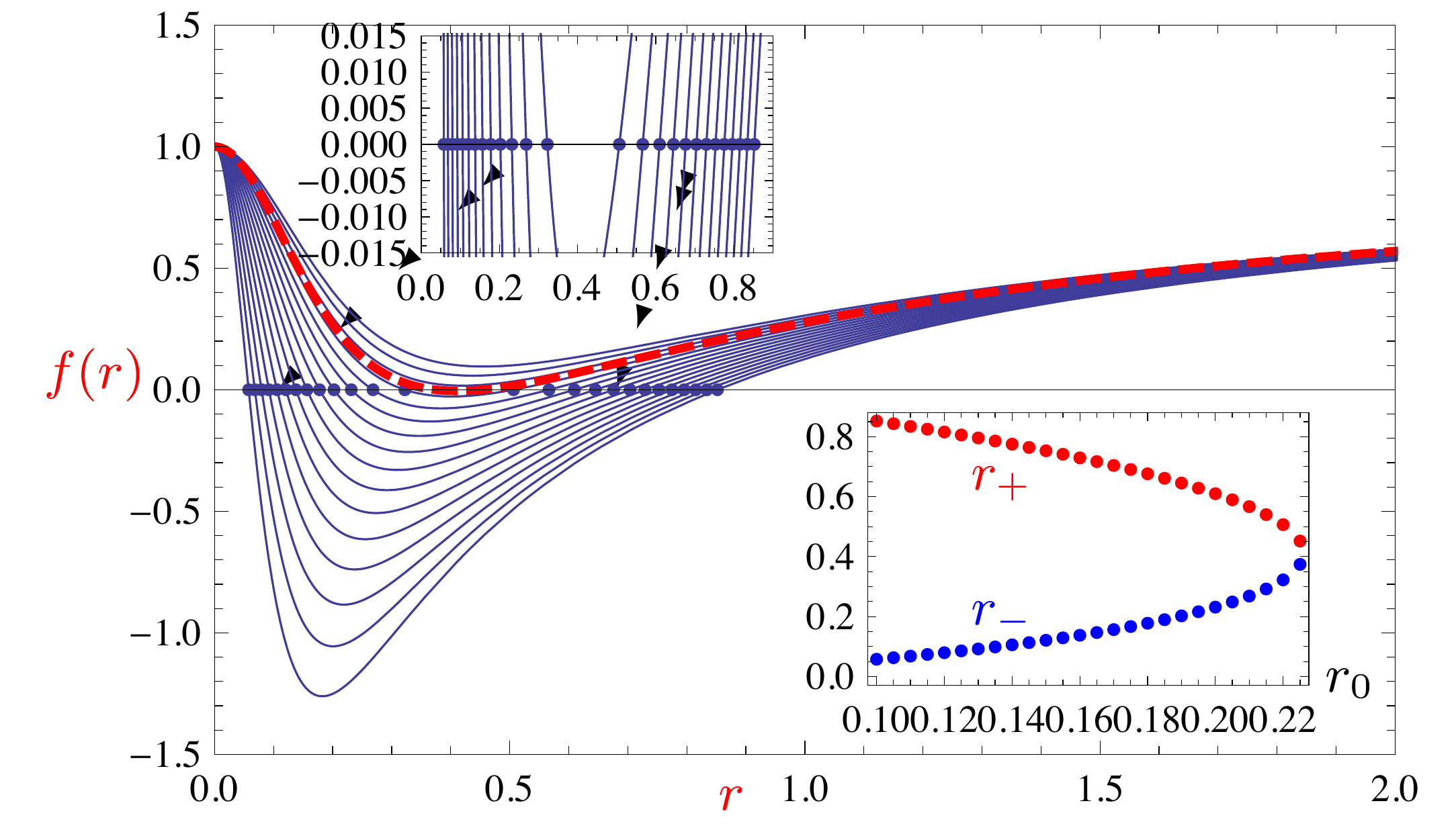}
\end{picture}
\caption{The behavior of the metric function
$f(r)$ given in (\ref{dymnf}) for different values of the parameter
$r_0$. The thick dashed line refers to the case where the two horizons
overlap. The bottom-right inset represents the values of the outer
(inner) horizon $r_+$ ($r_-$) {\it
vs} the parameter $r_0$. The mass $m$ has been normalized to unity and
$r_0$ varied up to its critical value above which the horizons
disappear and the geometry is regular.}
\label{fighor2}
\end{figure}

The basic properties of the Lagrangian
$\mathscr{L}\left(F\right)$ leading to electrically charged
configurations have been discussed originally in Bronnikov
\cite{bronnikov,berej} and used by Dymnikova \cite{dymn2} to show that
regular electrically charged solutions compatible with the weak energy
condition require a de Sitter center and to construct a prototypical
example. Specifically, Dymnikova's solution is obtained from a
nonlinear electrodynamic theory with a Hamiltonian-like function (see
\cite{dymn2} for details) of the form $\mathscr{H} = P \left(1 +\alpha \sqrt{-P}\right)^{-2}$, cwhere $P_{\mu\nu} = \mathscr{L}\left(F\right) F_{\mu\nu}$
($P=P_{\mu\nu}P^{\mu\nu}$) leading to a solution of the above form
with
\bea
f(r) = 1-{4m\over \pi r}\left(\mbox{tan}^{-1}{r\over r_0} -
{r r_0 \over r^2+r_0^2}\right)~.
\label{dymnf}
\eea
The parameter $r_0$ in the solution (\ref{dymnf}) is a length scale
related to the total mass $m$ and the charge $q$ by the relation
$r_0=\pi q^2/ (8m)$. The above solution reproduces asymptotically, for
$r\rightarrow \infty$, the Reissner-Nordstr\"om  behavior, while
near the center, for $r\ll r_0$ the solution approximates to de
Sitter. Depending on the value of $r_0$, the above regular solution
may present two distinct, one, or no horizons (see
Fig.~\ref{fighor2}).
\vspace{-0.3cm}
\vspace{-0.3cm}
Another electrically charged solution discussed in the literature is
the one presented by Ayon-Beato and Garc\'ia in
Ref.~\cite{ayonbeato}. It takes the form (\ref{rbhm}) with $f$ given
by
\bea
f= 1-{2mr^2\over (r^2+q^2)^{3/2}}+{q^2r^2\over (r^2+q^2)^2}~,
\label{lapseayon}
\eea
and it is obtained from a nonlinear electrodynamics with Lagrangian density
$
{\mathscr L} = {X^2\over -2q^2}{1-8X-3X^2 \over \left(1-X\right)^4}
-{3m\over 2 q^3 }{X^{5/2}\left(3-2X\right) \over \left(1-X\right)^{7/2}}~,
$
where $X=\sqrt{-2q^2F}$, and $m$ and $q$ are associated with mass and
charge, respectively. The above solution is asymptotically flat and
behaves like Reissner-Nordstr\"om, at leading order, asymptotically.
Depending on
the values of the charge and mass, as Dymnikova's solution
(\ref{dymnf}), it has two distinct inner and event horizons, for
values of the charge smaller than a critical value $q_c$, two
degenerate horizons for $q=q_c$, and becomes a globally regular
geometry for $q>q_c$. This solution, as discussed in
Ref.~\cite{bronnikov}, presents a problem related to the presence
of cusps in the electromagnetic Lagrangian. As pointed out in
Ref.~\cite{bronnikov}, such a solution should be taken with care, and
here we will use it as a useful working example and to check our numerics
against known results.

Amongst the various models, the one of Ref.~\cite{fabris} is slightly more
elaborate and the solution arises from the system of gravity coupled
to a phantom scalar field.

All the solutions considered in this paper are summarized in Table~I.
\begin{table}[ht]
\begin{tabular}{l|l}
Lapse function & Ref. \\ \hline
$f(r) = 1-\frac{2mr^2}{(r^2+\alpha^2)^{3/2}}$ &
\cite{bardeen,AyonBeato:2000zs}\\
$f=1-{2mr^2\over r^3+2\alpha^2}$ & \cite{sh} \\
$f=1-{2m\over r}\left(1- \mbox{tanh}{r_0\over r}\right)$ &
\cite{bronnikov,berej} \\
$f=1-{4m\over \pi r}\left(\mbox{tan}^{-1}{r\over r_0}-
{rr_0\over r^2+r_0^2}\right)$ & \cite{dymn2} \\
$f= 1-{2mr^2\over (r^2+q^2)^{3/2}}+{q^2r^2\over (r^2+q^2)^2}$
&\cite{ayonbeato} \\
$f=1+{cr^2\over b^2}+{\rho_0r^2\over b^3}
\left({b\sqrt{r^2-b^2}\over r^2} + \mbox{tan}^{-1}
{\sqrt{r^2-b^2}\over b} \right)$
&\cite{fabris} \\
\end{tabular}
\caption{The models of Refs. [1–8] were constructed in such a
   way to have at least Reissner-Nordstr\"om asymptotics (i.e., order of
   $1+ 1/r + 1/r^\alpha$ with $\alpha\geq2$, see Ref. [1,7]) or at
   least Reissner-Nordstr\"om-de Sitter asymptotics (i.e., order of
   $r^2+ 1/r + 1/r^\alpha$ with $\alpha\geq2$, see Ref. [8] for
   the case $\alpha=3$).
The solution \cite{bardeen,AyonBeato:2000zs}
can be put in a nonlinear electrodynamics framework.
The solution given in \cite{sh} is a minimal regular
solution with the parameter $m$ measuring the mass and the parameter
$\alpha$ measuring the deviation of the nonsingular solution from
Schwarzschild, which is reproduced for $\alpha=0$.  The models of
Refs.~\cite{bronnikov,berej,dymn2,ayonbeato} were constructed in the
context of nonlinear electrodynamics coupled to gravity using
different functional forms for $\mathscr L$.  The parameter $r_0$ in
\cite{bronnikov,berej,dymn2} is a length scale related to the electric
charge and $q$ in \cite{ayonbeato} is the electric charge itself.
The solution proposed in \cite{fabris} was constructed coupling gravity to
a phantom scalar field with a potential $U(b,c,\rho_0)$ where the
parameters $b,~c$, and $\rho_0$ characterize the features of the
potential. All the details can be found in the original references.}
\end{table}
\vspace{-0.8cm}
\section{Quasinormal Frequencies}

The computation of QNMs and their QNFs has been at the center
of great attention both for the astrophysical relevance in relation to
gravitational wave observation and for its formal importance.
Extensive computations of QNFs for various black hole geometries are
now available and most results have been reviewed
\cite{SchutzWill,iw,iyer2,leaver,Hod:1997mt,Konoplya:2002ky,kokkotasberti,neitz,Konoplya:2011qq,Motl:2002hd,Motl:2003cd,Nollert:1999ji,Kokkotas:1999bd,Berti:2009kk}.

Here we extend the computation of the QNFs to regular black hole
geometries; see also \cite{fernando,bronkonzid}.
Although we will consider the
specific solutions listed in Table I, it is worth noticing that the
QNFs for uncharged scalar perturbations, as well as for tensor
perturbations, depend on the specific form of the matter Lagrangian
used to obtain the background solution only through the explicit
functional form of the lapse function $f$. The same is also true for
the case of charged scalar and vector perturbations, as long as the
weak field limit of the nonlinear electrodynamics reproduces the
standard Maxwell's theory.

\vspace{-0.4cm}

\subsection{Neutral scalar perturbations}
The computation of the QNFs for regular black hole geometries proceeds
as in the well known cases, and this section will be devoted to those
arising from neutral scalar field perturbations. The equation for
these perturbations takes the usual form,
\bea
{1\over \sqrt{g}}\partial_\mu \left(\sqrt{g}g^{\mu\nu}\partial_\nu
\right)\phi =0~,
\label{sceq}
\eea
where $g$ is the determinant of the metric tensor $g_{\mu\nu}$
given in (\ref{rbhm}) and the lapse function $f$ depends on the model
considered.
Equation (\ref{sceq}) can be separated
by decomposing the scalar perturbation into appropriate harmonics,
\bea
\phi = r^{-1}\sum_{lm} e^{-i \omega t}\varphi_{lm}(r)Y_l^m(\Omega)~,
\eea
and by introducing the tortoise coordinate,
\bea
dx = dr/ f~.
\nonumber
\eea
One then can
rewrite the radial part of (\ref{sceq}) in a Schr\"odinger
form,
\bea
\left[-{d^2\over dx^2} + V(x) - \omega^2\right] \varphi(x)=0~,
\eea
with $V$ given by
\bea
V =
f\left({\ell(\ell+1)\over r^2} + {f'\over r}\right)~,
\label{rbhp}
\eea
and where the indices $l,m$ have been suppressed from
$\varphi_{lm}$.
The boundary conditions, appropriate for the computation of the
QNFs, must
force the solution near the event horizon, $x=-\infty$ (at infinity,
$x=+\infty$), not to generate outgoing (ingoing) waves. These can be
written as
\bea
\varphi &\sim& e^{- i \omega x}~,~~~ x\rightarrow -\infty~,\nonumber\\
\varphi &\sim& e^{+ i \omega x}~,~~~ x\rightarrow +\infty~.\nonumber
\eea
In the present case, most available methods to compute QNFs would
work.

Here, we will adopt the most direct way that makes use of WKB
approximation, originally discussed in Ref.~\cite{iw}, and that does
not require any special modification.
Results for the quasinormal frequencies for neutral scalar
perturbations are tabulated, in Appendix A, in Table II for the model
given in Refs.~\cite{bardeen,AyonBeato:2000zs}, in Table III for the
model given in Ref.~\cite{sh}, in Table IV for the model given in
Refs.~\cite{bronnikov,berej}, in Table V for the model given in
Ref.~\cite{dymn2}, in Table VI for the model given in
Ref.~\cite{ayonbeato}, and in Table VII for the model given in
Ref.~\cite{fabris}.
The tables list the values of the frequencies at third order in the WKB
approximation, as results to this order allow a direct comparison with
the results presented in
\cite{iyer2,Konoplya:2002ky}. One can go
up to sixth order in the WKB approximation and this allows to
check the convergence of the approximation and how the accuracy
improves going to higher order.

Figure 2 illustrates the real and imaginary parts of the frequencies
for the solution of Dymnikova
\cite{dymn2} and for sample values of the parameters,
multipole number $l$, and overtone number $n$, up to sixth order WKB,
thus allowing to test the convergence of the approximation.
\begin{figure}[ht]
\begin{center}
\unitlength=1mm
\begin{picture}(75,50)
  \includegraphics[height=5cm]{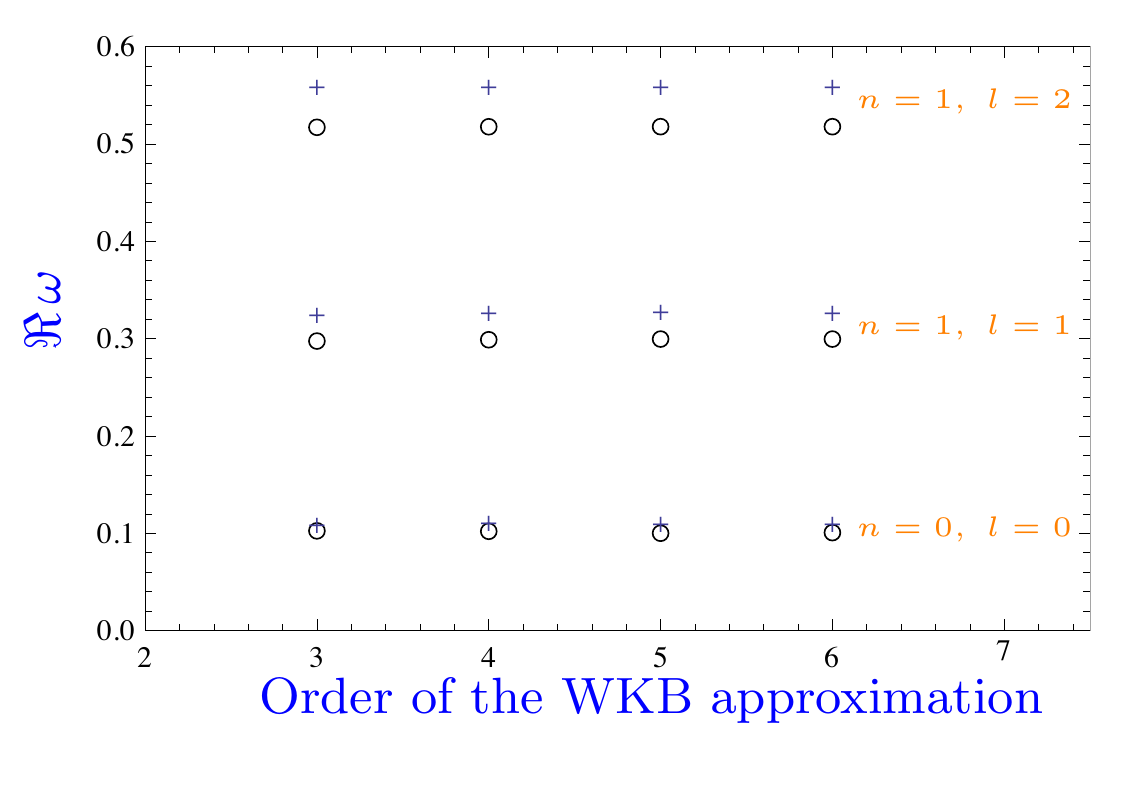}
\end{picture}
\end{center}
\label{fig1}
\end{figure}
\begin{figure}[ht]
\begin{center}
\unitlength=1mm
\begin{picture}(77,40)
  \includegraphics[height=5cm]{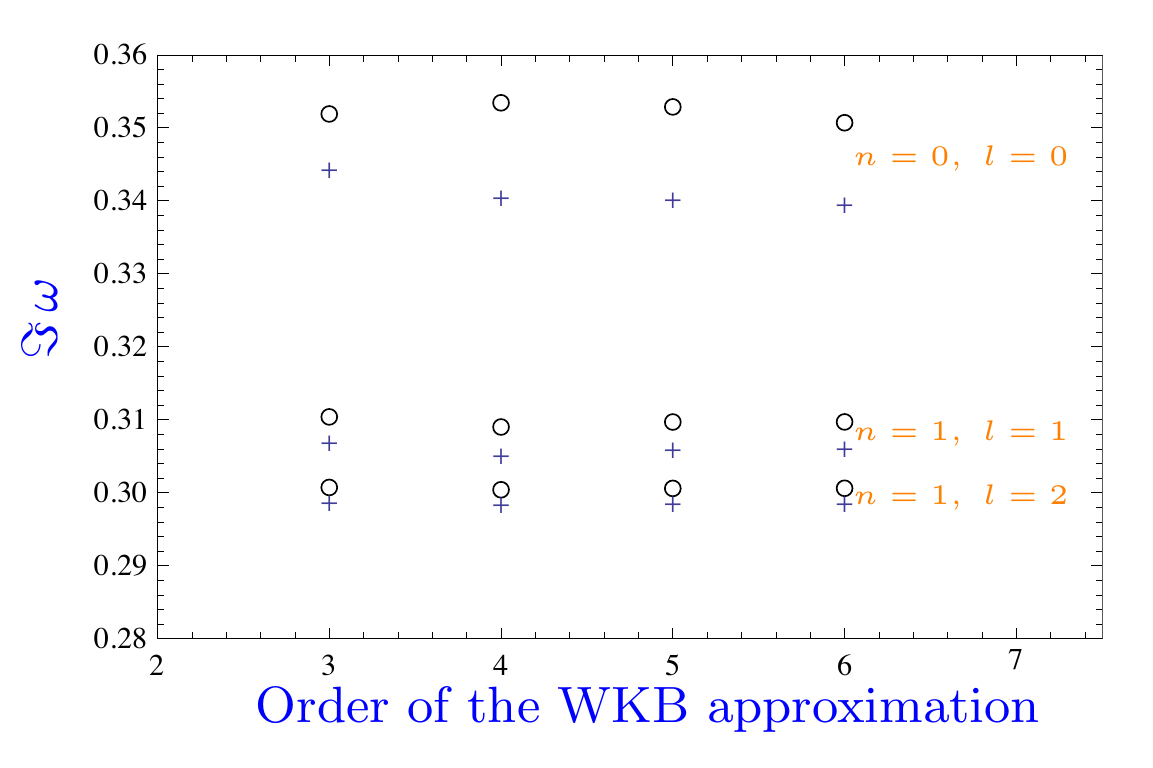}
\end{picture}
\end{center}
\caption{The real (upper panel) and imaginary
(lower panel) parts of the QNFs
from neutral scalar perturbations for sample values of the overtone and
multipole numbers for the model of Dymnikova \cite{dymn2} when higher order
terms in the WKB approximation are included in the computation. The
symbols $\circ$ ($+$) refer to $r_0=0.2$ ($r_0=0.3$) and
we have normalized $m$ to unity.
The overtone and multipole numbers are indicated in the figure.}
\label{fig1b}
\end{figure}

Because of its connection with a
possible black hole area quantization law, it seems
interesting to study the behavior of the QNFs in the limit of the large
imaginary part. The problem is rather similar to the case of charged
black hole solutions studied, for example, in Ref.~\cite{Motl:2003cd}. 
This comes as no surprise, since the regular black hole solutions
   considered here have, at leading order, an asymptotic structure at
   least of the Reissner-Nordstr\"om type or Reissner-Nordstr\"om-de
   Sitter type, i.e., order of $1+ 1/r + 1/r^\alpha$ with
   $\alpha\geq2$ (see [1,7]), or order of $r^2+ 1/r + 1/r^\alpha$ with
   $\alpha\geq2$ (see [8] for the case $\alpha=3$), respectively.
Although formally the process
of monodromy matching can be performed in the same way, one can
see that in the limit $r\rightarrow 0$, the leading term in the
potential (\ref{rbhp}) behaves as $l(l+1)r^{-2}$, rather than
$r^{-6}$. This suggests that the QNFs, in the limit of the large imaginary
part will depend on the multipole number $l$. Following the same steps
as in \cite{Motl:2003cd}, one finds that asymptotically the
frequencies will satisfy the following relation:
\bea
e^{\beta \omega} &=& - ( 1 + 2 \cos( \pi \sqrt{1+l(l+1)} ) )
\nonumber\\ &&- e^{\beta_- \omega} ( 2 + 2\cos ( \pi \sqrt{1+l(l+1)}) )~,
\eea
where $\beta$ and $\beta_-$ are the inverse temperature at the outer
and inner horizons. The above formula suggests that the real part of
the QNFs does not asymptote to a constant value, nor display any
periodic behavior. It may be interesting to study what happens in
the extremal case. If the analogy with the Reissner-Nordstr\"om black hole is valid also in
this case, then, as the charges approach the extremal value, the
asymptotic value of the real part of the QNFs would spiral toward the
asymptotic Schwarzschild QNFs \cite{kokkotasberti,neitz}. To test this,
however, a more suitable numerical approach is necessary and this is
left for future work.


\vspace{-0.3cm}

\subsection{Charged scalar perturbations}

The computation of QNFs can be extended to charged scalar
perturbations. The problem is similar to the Reissner-Nordstr\"om case
treated in Refs.~\cite{leaver,Hod:1997mt,kokkotasberti,Konoplya:2002ky}
where QNFs associated with charged scalar perturbations were
computed. In particular, Ref.~\cite{Konoplya:2002ky} followed the
analysis of Ref.~\cite{Hod:1997mt} that studied the evolution of
scalar perturbations around collapsing charged black holes, indicating
that scalar perturbations are radiated away, at a slower rate than the
neutral ones, and follow an inverse power-law behavior at the future
outer horizon, while displaying a decaying behavior accompanied by
oscillations at the future outer horizon.

The general equation for the
(complex) scalar perturbations can be written formally in the same way
as \cite{Hod:1997mt},
\bea
\phi_{;ab}g^{ab}-i e A_a g^{ab}(2\phi_{;b} - ieA_b\phi) -
ieA_{a;b}g^{ab}\phi=0,~~~\label{eom}
\eea
with $e$ being the (constant) charge of the scalar field. Since we are
concerned with perturbations, the electromagnetic potential for the
black hole is determined by the ordinary Maxwell's theory and can be
written, up to an additive integration constant $c$, as
\bea
A_{\mu}dx^{\mu}=-{q\over r}dt+c~.
\nonumber
\eea
The integration constant is to be determined by regularity at the
horizon for the gauge field. In fact, as far as the evaluation of the
QNFs is concerned, its precise value is unessential since it would
only produce a uniform shift in the real part of the frequencies. In
the following, we will set $c=0$. Equation (\ref{eom}) can be separated by
decomposing the scalar perturbation into appropriate harmonics,
and using tortoise coordinates
it is possible to rewrite the above equation in Schr\"odinger form,
\bea
\left[-{d^2\over dx^2} + V(x) - \left(\omega+eq/r\right)^2\right]
\varphi(x)=0~,
\eea
where $V$ is given by (\ref{rbhp}).
The boundary conditions relevant for the QNFs computation are
\bea
\varphi &\sim& e^{- i \left(\omega+eq/r_+\right) x}~,~~~ x\rightarrow -
\infty~,\nonumber\\
\varphi &\sim& e^{+ i \omega x}~,~~~ x\rightarrow +\infty~.\nonumber
\eea
The WKB method works in this case too and the computation traces over
the Reissner-Nordstr\"om case, with the difference being the lapse
function $f(r)$, requiring a small adaptation in the numerics. In the
case of charged perturbation some more work is necessary, but the same
approach of analytic continuation used in Ref.~\cite{Konoplya:2002ky}
can be adopted here: first fix all the parameters including the
multipole and overtone numbers, then maximize the potential as a
function of the radial distance, and finally find the value of omega
that satisfies the WKB relation for the QNFs as a numerical function
analytically continued into the complex domain (the potential is
generally a complex function).
The results for the QNFs for charged scalar perturbations are listed
in Appendix B, in Table VIII for the model given in
Refs.~\cite{bardeen,AyonBeato:2000zs}, in Table IX for the model in
Ref.~\cite{dymn}, and in Table X for the model in Ref.~\cite{ayonbeato}.

\begin{figure}[ht]
\begin{center}
\unitlength=1mm
\begin{picture}(140,60)
  \includegraphics[height=5.8cm]{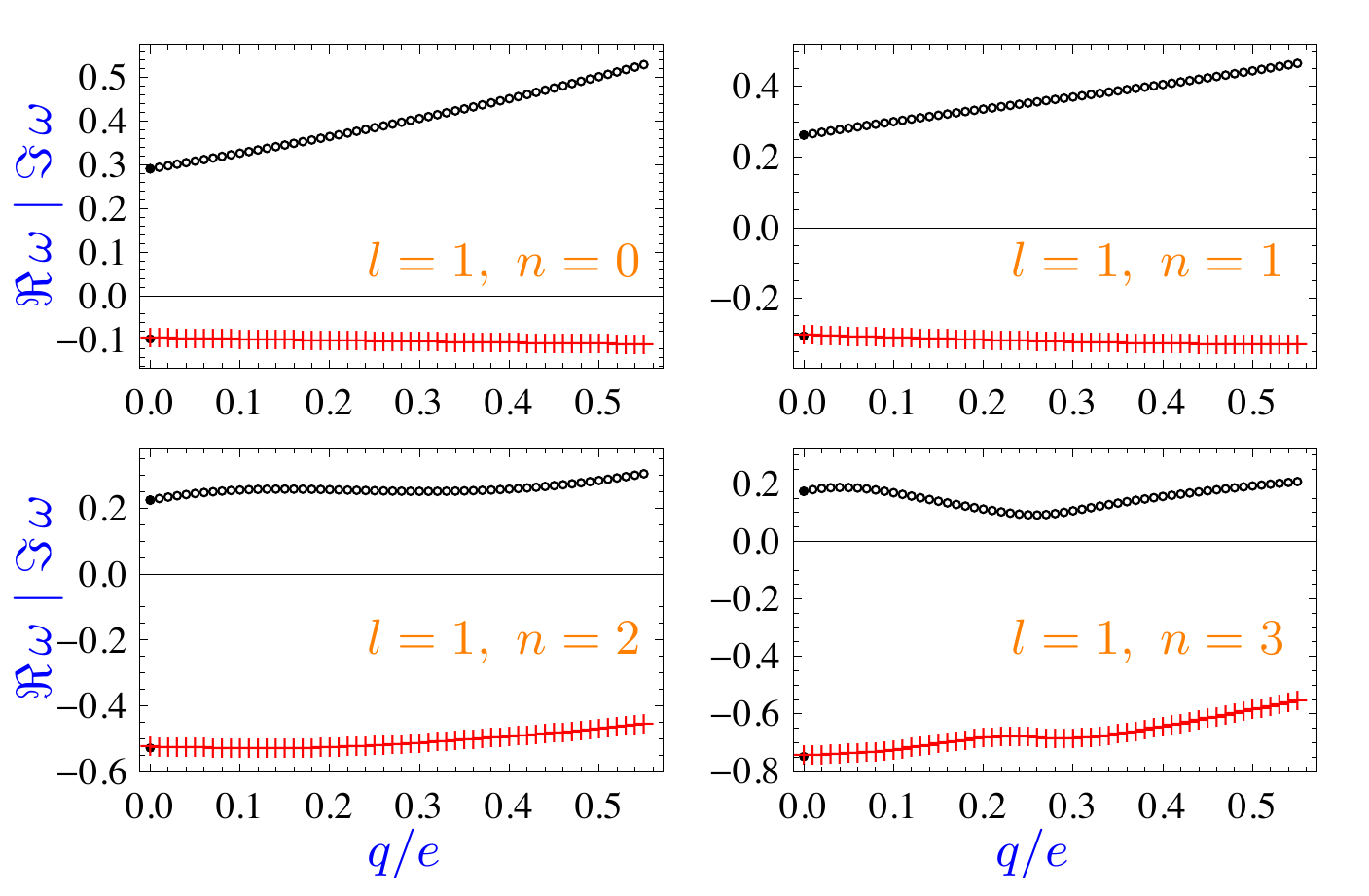}
\end{picture}
\end{center}
\caption{The real ($\circ$) and imaginary ($+$)
parts of the QNFs with $l=1$
from charged scalar perturbations for sample values
of the overtone and multipole numbers (as indicated in the panels) for
the model of Dymnikova \cite{dymn2}. The dark circles represent the
values of the QNFs for the Schwarzschild case that are continuously
connected with the values for the Dymnikova solution in the limit of
zero charge ($q=0$).}
\label{figdymch1}
\end{figure}

\begin{figure}[ht]
\begin{center}
\unitlength=1mm
\begin{picture}(100,60)
  \includegraphics[height=5.8cm]{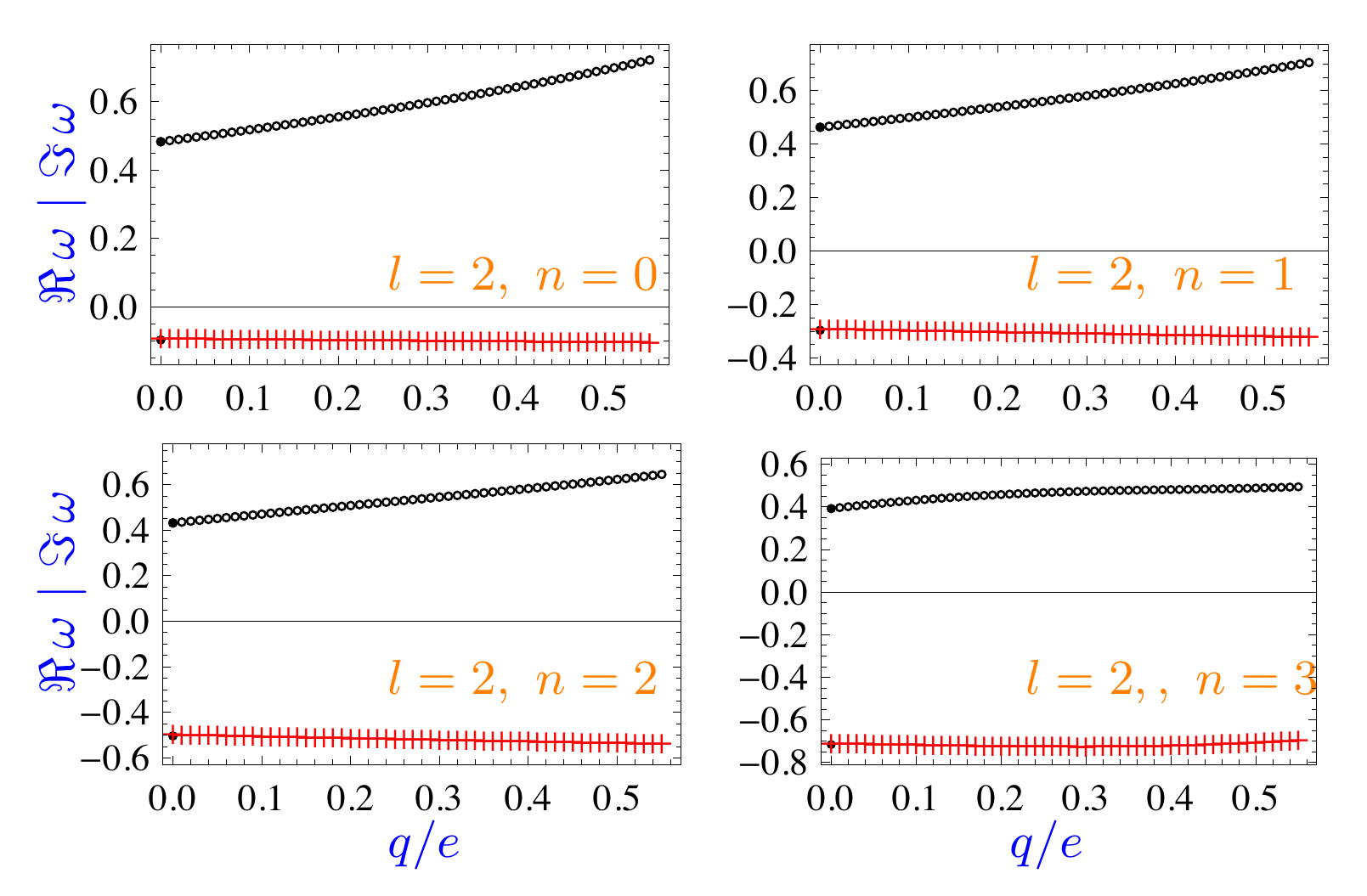}
\end{picture}
\end{center}
\caption{Same as Fig.~\ref{figdymch1} with $l=2$.}
\label{figdymch2}
\end{figure}

\begin{figure}[ht]
\unitlength=1mm
\begin{picture}(85,50)
  \includegraphics[height=5.cm]{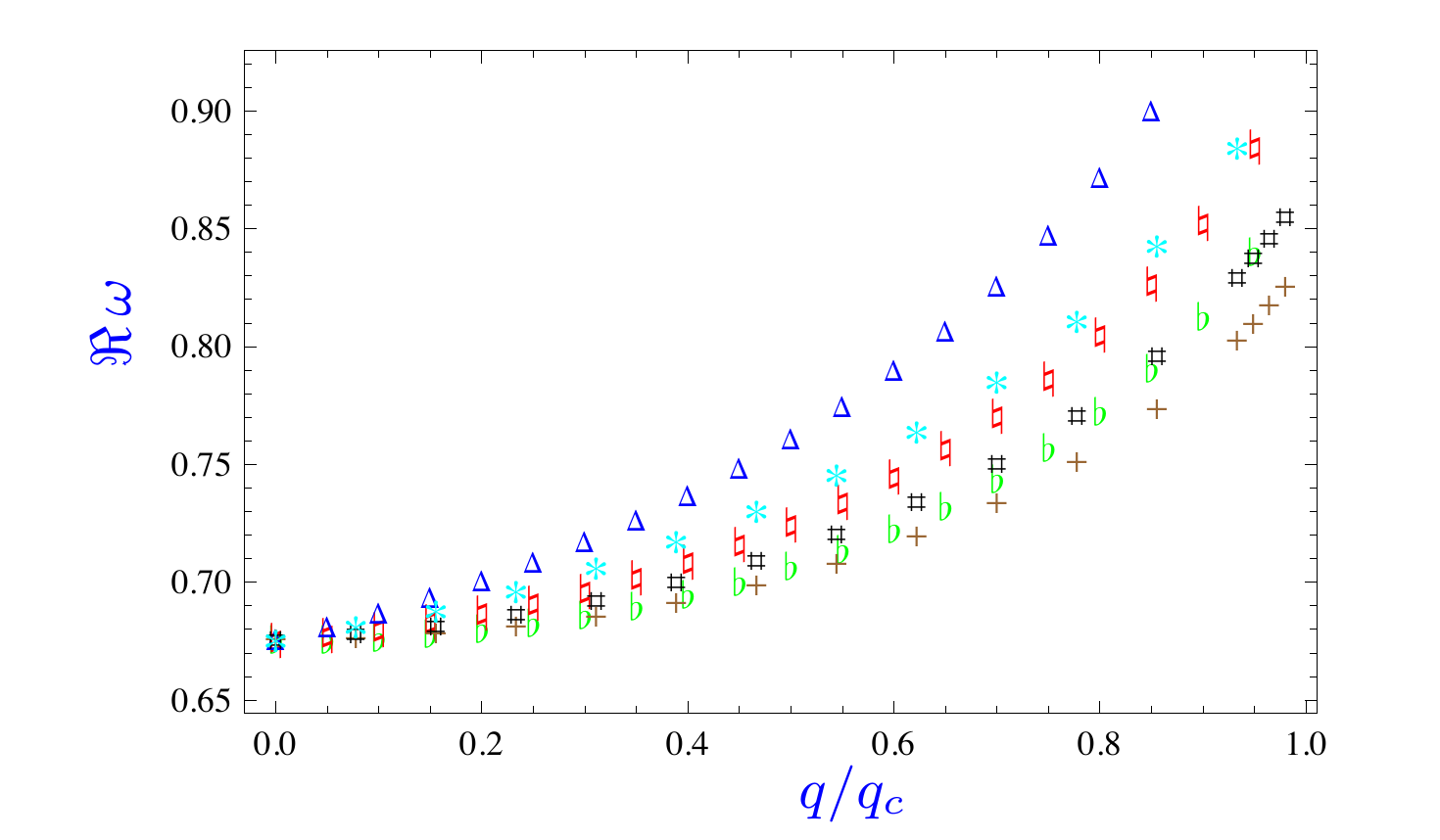}
\end{picture}
\caption{The behavior of the real part of the
QNFs for $l=3$ and $n=0$ in the Reissner-Nordstr\"om case and in the
regular black hole case for the model given in \cite{ayonbeato},
see Eq.~(\ref{lapseayon}). The symbols refer to
$q=0\,(\small{+}),~0.1\,(\small{\sharp}),~0.3\,(\small{*})$ for the
regular case, and
$q=0\,(\small{\flat}),~0.1\,(\small{\natural}),~0.3\,(\small{\Delta})$
for the Reissner-Nordstr\"om case.}
\label{fig1c}
\end{figure}

\begin{figure}[ht]
\unitlength=1mm
\begin{picture}(95,50)
  \includegraphics[height=5.cm]{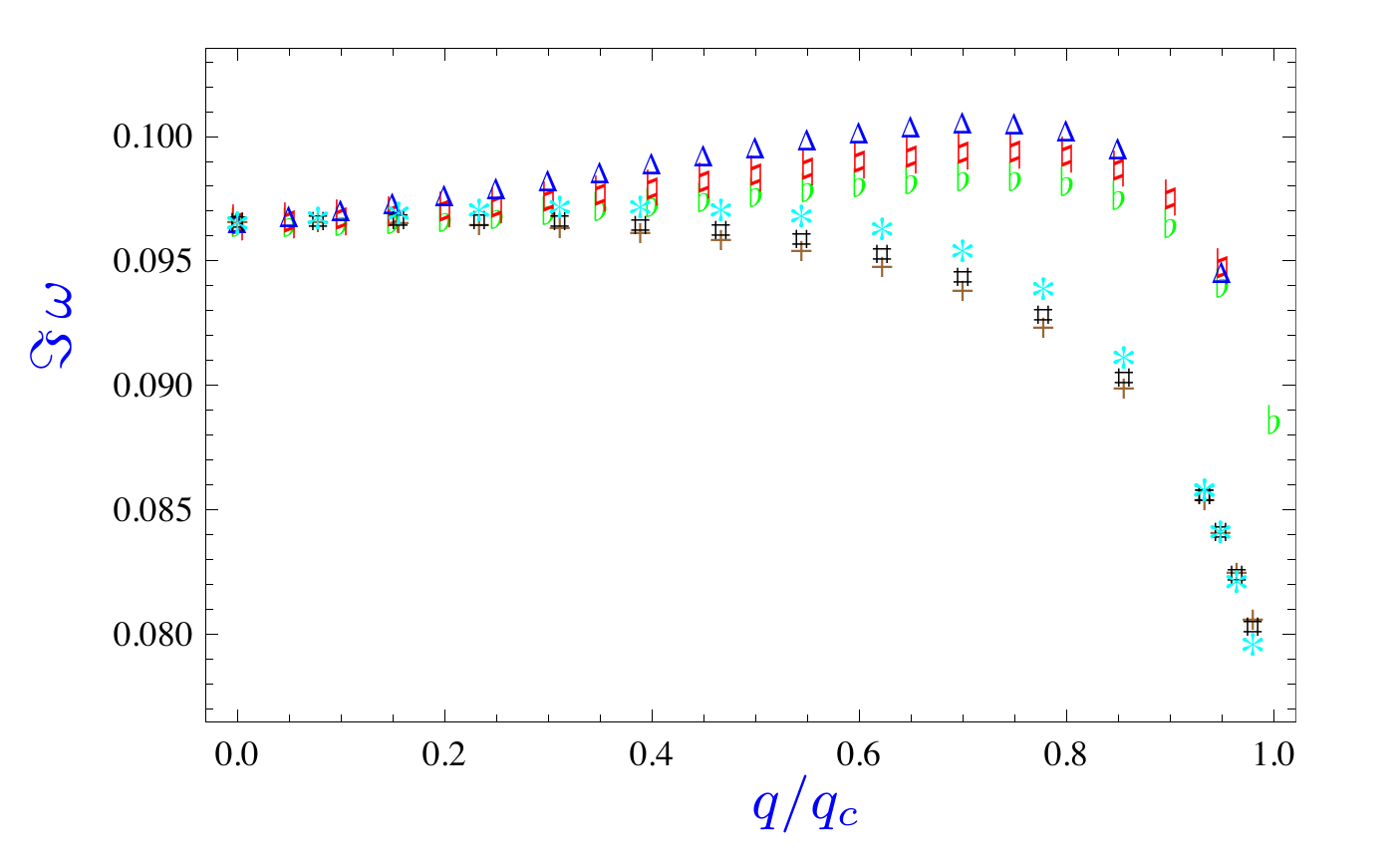}
\end{picture}
\caption{Same as Fig.~\ref{fig1c} for the imaginary part of the QNFs.}
\label{fig3s}
\end{figure}

Figures \ref{figdymch1} and \ref{figdymch2} illustrate the behavior of the
QNFs with respect to the charge $q$ for the solution of \cite{dymn2}. It
is possible to notice that for vanishing $q$ the solution reproduces
the Schwarschild geometry. In this limit, scalar perturbations become
uncharged and the value of the QNFs must be smoothly connected with
the QNFs for uncharged scalar perturbations on Schwarzschild black
holes, indicated by the black dots in
Figs.~\ref{figdymch1}-\ref{figdymch2}.
Figures \ref{fig1c} and \ref{fig3s} illustrate the QNFs for the $l=3$ and
$n=0$ mode for different values of the charge of the perturbation in
the Reissner-Nordstr\"om case and in the regular black hole case. The
choice of these specific values for the multipole and overtone numbers
was motivated by the desire to compare our results with those of
Ref.~\cite{Konoplya:2002ky}.
One notices (upper panel of Fig.~\ref{fig3s}) that the real part of the
QNFs follows a behavior analogous to the Reissner-Nordstr\"om case;
i.e.,  $\Re\,\omega$ grows with the charge and it is larger for
charged perturbations than for neutral ones.
This is a natural
consequence of the fact that the regular black hole geometries
considered asymptote, at leading order, to Reissner-Nordstr\"om. This
indicates that, in the regular case as well, the late time behavior of
the quasinormal ringing will be dominated by the neutral scalar
perturbations.  The lower panel illustrates the imaginary part of the
QNFs for Reissner-Nordstr\"om and for the regular black hole solution
(\ref{rbhm}). While for small values of the charge $q$, i.e., in
the region where the deviations from the two geometries are small, the
values of $\Im\,\omega$ are basically superposed, the discrepancy
increases with the charge signifying that in the large $q$ region the
exponential damping of these modes will occur less prominently than in
the Reissner-Nordstr\"om case. The coincidence in the imaginary parts
of the QNFs is observed in the regular case as well, but the feature
of the profile, namely the fact that $\Im\,\omega$ undergoes a
transient increase before dropping for large values of the charge, is
much less prominent in the regular case, where the increase in
$\Im\,\omega$ for moderate values of $q$ is small.

\vspace{-0.3cm}
\section{Conclusions}
In this work we have computed the neutral and charged
scalar QNFs for regular black hole geometries.
The prototype example of such regular solutions is the one proposed by
Bardeen many years ago \cite{bardeen}
(see also \cite{AyonBeato:2000zs}), and several examples were
constructed following this original example. 
Such solutions are spherically symmetric, multihorizon, and singularity-free and display an asymptotic behavior analogous, at leading order, to at least Reissner-Nordstr\"om or Reissner-Nordstr\"om-de Sitter, while having a near-horizon behavior similar to Schwarzschild.
The WKB method can be implemented in
the present case, and it has been adopted to
compute the QNFs for the variety of models of
Refs.~\cite{bardeen,AyonBeato:2000zs,sh,bronnikov,berej,dymn2,fabris,ayonbeato}.
The results have been
tested in the limit of Schwarzschild and Reissner-Nordstr\"om black
holes reproducing the known results of
Refs.\cite{iyer2,Konoplya:2002ky} when the same order in the WKB
approximation is used and the convergence has been tested up to sixth
order in the WKB expansion. The QNMs of Bardeen's model have
also been studied in \cite{fernando,bronkonzid}.

Aside of the generalization to other types of perturbations, the use
of more sophisticated numerical methods, like those used in
Ref.~\cite{kokkotasberti}, may be useful to check the accuracy of the
WKB approximation as well we to test the asymptotic behavior of the
QNFs. These problems are left for future work.

Certainly, knowledge of how a black hole rings after being perturbed
can shed light on some fundamental aspects of quantum gravity. In
turn, knowledge of quantum gravity should provide us with a better
understanding of black holes and eventually suggest a possible
resolution of the singularity problem.  In fact, the presence of
singularities certainly signals a limitation of our understanding, if
not a breakdown, of general relativity. This pathological behavior is
usually believed to disappear in a full theory of quantum gravity that
would provide a consistent framework to test the well known
semiclassical arguments predicting the evaporation of black holes.
\vspace{0.3cm}

\acknowledgments
The support of the Funda\c{c}\~{a}o p\^{a}ra a Ci\^{e}ncia e a
Tecnologia of Portugal (Project PTDC/FIS/098962/2008) and of the
European Union Seventh Framework Programme (Grant Agreement
PCOFUND-GA-2009-246542) is gratefully acknowledged.  Thanks are due
R. Konoplya for correspondence that allowed to compare our results
with those of Ref.~\cite{Konoplya:2002ky}. We also wish extend our
acknowledgements to V. Cardoso and P. Pani for many useful discussions
and help in checking the numerical results presented in this paper and
to S. Ansoldi for earlier discussions.

\newpage
\appendix
\begin{widetext}
\section{Tables of QNFs for neutral scalar perturbations}
\begin{table}[ht]
\begin{tabular}{llccc}
 \hline
$l$ ~~~ ~~~& $n$ ~~~~~~& $q=0.1$~~~& $q=0.3$~~~& $q=0.6$\\
 \hline
0  & 0 & $0.1049-\imath~0.1149$ ~~~& $0.1070-\imath~0.1125$ ~~~& $0.1093-\imath~0.1015$\\
  & 1 & $0.0895-\imath~0.3541$ ~~~& $0.0921-\imath~0.3473$ ~~~& $0.0866-\imath~0.3219$\\
\hline
1   & 0 & $0.2916-\imath~0.0978$ ~~~& $0.2959-\imath~0.0967$ ~~~& $0.3132-\imath~0.0906$\\
   & 1 & $0.2629-\imath~0.3069$ ~~~& $0.2686-\imath~0.3028$ ~~~& $0.2886-\imath~0.2820$\\
   & 2 & $0.2245-\imath~0.5259$ ~~~& $0.2323-\imath~0.5187$ ~~~& $0.2548-\imath~0.4831$\\
   & 3 & $0.1750-\imath~0.7474$ ~~~& $0.1860-\imath~0.7368$ ~~~& $0.2125-\imath~0.6869$\\
\hline
2   & 0 & $0.4840-\imath~0.0966$ ~~~& $0.4909-\imath~0.0956$ ~~~& $0.5191-\imath~0.0901$\\
   & 1 & $0.4641-\imath~0.2954$ ~~~& $0.4721-\imath~0.2920$ ~~~& $0.5034-\imath~0.2740$\\
   & 2 & $0.4328-\imath~0.5027$ ~~~& $0.4424-\imath~0.4966$ ~~~& $0.4777-\imath~0.4645$\\
   & 3 & $0.3940-\imath~0.7148$ ~~~& $0.4058-\imath~0.7059$ ~~~& $0.4454-\imath~0.6595$\\
 \hline
\end{tabular}
\caption{QNFs for neutral scalar perturbations for $q=0.1,~0.3,~0.6$
and $m=1$ for the model of Refs.~\cite{bardeen,AyonBeato:2000zs}. The
model reproduces the original example constructed by Bardeen using a
nonlinear electromagnetic theory that displays, in the weak field
limit, a stronger behavior as compared to the ordinary Maxwell's
one. In this model the gravitational field can be interpreted as that
of a nonlinear magnetic monopole.}
\end{table}
\centerline{}
\begin{table}[ht]
\begin{tabular}{llccc}
 \hline
$l$ ~~~ ~~~& $n$ ~~~~~~& $\alpha=0.1$~~~& $\alpha=0.4$~~~& $\alpha=0.7$\\
 \hline
0  & 0 & $0.1046-\imath~0.1148$ ~~~& $0.1034-\imath~0.1093$ ~~~& $0.0923-\imath~0.0986$\\
  & 1 & $0.0891-\imath~0.3541$ ~~~& $0.0835-\imath~0.3424$ ~~~& $0.0562-\imath~0.3281$\\
\hline
1   & 0 & $0.2913-\imath~0.0978$ ~~~& $0.2946-\imath~0.0948$ ~~~& $0.3018-\imath~0.0854$\\
   & 1 & $0.2624-\imath~0.3068$ ~~~& $0.2655-\imath~0.2973$ ~~~& $0.2649-\imath~0.2691$\\
   & 2 & $0.2238-\imath~0.5258$ ~~~& $0.2264-\imath~0.5104$ ~~~& $0.2108-\imath~0.4677$\\
   & 3 & $0.1741-\imath~0.7473$ ~~~& $0.1764-\imath~0.7262$ ~~~& $0.1418-\imath~0.6727$\\
\hline
2   & 0 & $0.4835-\imath~0.0966$ ~~~& $0.4892-\imath~0.0940$ ~~~& $0.5037-\imath~0.0855$\\
   & 1 & $0.4636-\imath~0.2953$ ~~~& $0.4700-\imath~0.2870$ ~~~& $0.4823-\imath~0.2601$\\
   & 2 & $0.4321-\imath~0.5025$ ~~~& $0.4392-\imath~0.4881$ ~~~& $0.4445-\imath~0.4422$\\
   & 3 & $0.3931-\imath~0.7146$ ~~~& $0.4009-\imath~0.6942$ ~~~& $0.3946-\imath~0.6310$\\
 \hline
\end{tabular}
\caption{QNFs for neutral scalar perturbations $\alpha=0.1,~0.4,~0.6$
and $m=1$ for the solution of Refs.~\cite{sh}.}
\end{table}
\begin{table}[ht]
\begin{tabular}{llccc}
 \hline
$l$ ~~~ ~~~& $n$ ~~~~~~& $r_0=0.1$~~~& $r_0=0.3$~~~& $r_0=0.4$\\
 \hline
0  & 0 & $0.1091-\imath~0.1151$ ~~~& $0.1189-\imath~0.1119$ ~~~& $0.1200-\imath~0.1065$\\
  & 1 & $0.0940-\imath~0.3550$ ~~~& $0.1021-\imath~0.3482$ ~~~& $0.0940-\imath~0.3401$\\
\hline
1   & 0 & $0.3019-\imath~0.0988$ ~~~& $0.3303-\imath~0.0990$ ~~~& $0.3503-\imath~0.0962$\\
   & 1 & $0.2742-\imath~0.3094$ ~~~& $0.3058-\imath~0.3080$ ~~~& $0.3252-\imath~0.2980$\\
   & 2 & $0.2374-\imath~0.5297$ ~~~& $0.2729-\imath~0.5264$ ~~~& $0.2897-\imath~0.5098$\\
   & 3 & $0.1902-\imath~0.7523$ ~~~& $0.2319-\imath~0.7472$ ~~~& $0.2451-\imath~0.7249$\\
\hline
2   & 0 & $0.5007-\imath~0.0977$ ~~~& $0.5472-\imath~0.0982$ ~~~& $0.5810-\imath~0.0957$\\
   & 1 & $0.4815-\imath~0.2984$ ~~~& $0.5306-\imath~0.2990$ ~~~& $0.5652-\imath~0.2904$\\
   & 2 & $0.4514-\imath~0.5074$ ~~~& $0.5041-\imath~0.5071$ ~~~& $0.5387-\imath~0.4916$\\
   & 3 & $0.4141-\imath~0.7212$ ~~~& $0.4715-\imath~0.7199$ ~~~& $0.5051-\imath~0.6976$\\
 \hline
\end{tabular}
\caption{QNFs for neutral scalar perturbations for $r_0=0.1,~0.3,~0.4$
for the solution of Refs.~\cite{bronnikov,berej}. This solution is
obtained in the context of gravity plus nonlinear electrodynamics
\cite{bronnikov}. The solution has also been extended to the case when
higher order curvature corrections are included in the gravitational
action Ref.~\cite{berej}. The parameter $r_0$ is a length scale
related to the charge.}
\end{table}
\begin{table}[ht]
\begin{tabular}{llccc}
 \hline
$l$ ~~~ ~~~& $n$ ~~~~~~& $r_0=0.1$~~~& $r_0=0.3$~~~& $r_0=0.4$\\
 \hline
0  & 0 & $0.1104-\imath~0.1150$ ~~~& $0.1241-\imath~0.1104$ ~~~& $0.1275-\imath~0.1030$\\
  & 1 & $0.0954-\imath~0.3548$ ~~~& $0.1073-\imath~0.3440$ ~~~& $0.0993-\imath~0.3314$\\
\hline
1   & 0 & $0.3051-\imath~0.0991$ ~~~& $0.3452-\imath~0.0989$ ~~~& $0.3771-\imath~0.0940$\\
   & 1 & $0.2779-\imath~0.3098$ ~~~& $0.3227-\imath~0.3065$ ~~~& $0.3549-\imath~0.2896$\\
   & 2 & $0.2416-\imath~0.5303$ ~~~& $0.2924-\imath~0.5232$ ~~~& $0.3220-\imath~0.4940$\\
   & 3 & $0.1953-\imath~0.7531$ ~~~& $0.2550-\imath~0.7422$ ~~~& $0.2806-\imath~0.7021$\\
\hline
2   & 0 & $0.5060-\imath~0.0980$ ~~~& $0.5716-\imath~0.0982$ ~~~& $0.6254-\imath~0.0936$\\
   & 1 & $0.4871-\imath~0.2990$ ~~~& $0.5564-\imath~0.2984$ ~~~& $0.6119-\imath~0.2834$\\
   & 2 & $0.4573-\imath~0.5083$ ~~~& $0.5321-\imath~0.5054$ ~~~& $0.5885-\imath~0.4783$\\
   & 3 & $0.4207-\imath~0.7224$ ~~~& $0.5023-\imath~0.7170$ ~~~& $0.5583-\imath~0.6775$\\
 \hline
\end{tabular}
\caption{QNFs for neutral scalar perturbations for $r_0=0.1,~0.3,~0.4$
and $m=1$ for the model of Ref.~\cite{dymn2}.}
\end{table}
\begin{table}[ht]
\begin{tabular}{llccc}
 \hline
$l$ ~~~ ~~~& $n$ ~~~~&$q=0$~~~& $q=0.3$~~~& $q=0.6$\\
 \hline
0  & 0 & $0.1046-\imath~0.1152$ ~~~& $0.1089-\imath~0.1123$ ~~~& $0.1118-\imath~0.0970$\\
   & 1 & $0.0892-\imath~0.3549$ ~~~& $0.0940-\imath~0.3470$ ~~~& $0.0804-\imath~0.3177$\\
\hline
1   & 0 & $0.2911-\imath~0.0980$ ~~~& $0.3009-\imath~0.0970$ ~~~& $0.3424-\imath~0.0861$\\
   & 1 & $0.2622-\imath~0.3074$ ~~~& $0.2742-\imath~0.3034$ ~~~& $0.3153-\imath~0.2667$\\
   & 2 & $0.2235-\imath~0.5268$ ~~~& $0.2386-\imath~0.5195$ ~~~& $0.2737-\imath~0.4581$\\
   & 3 & $0.1737-\imath~0.7486$ ~~~& $0.1934-\imath~0.7378$ ~~~& $0.2207-\imath~0.6545$\\
\hline
2   & 0 & $0.4832-\imath~0.0968$ ~~~& $0.4990-\imath~0.0960$ ~~~& $0.5689-\imath~0.0858$\\
   & 1 & $0.4631-\imath~0.2958$ ~~~& $0.4806-\imath~0.2929$ ~~~& $0.5531-\imath~0.2599$\\
   & 2 & $0.4316-\imath~0.5034$ ~~~& $0.4515-\imath~0.4979$ ~~~& $0.5248-\imath~0.4393$\\
   & 3 & $0.3925-\imath~0.7158$ ~~~& $0.4158-\imath~0.7076$ ~~~& $0.4871-\imath~0.6236$\\
 \hline
\end{tabular}
\caption{QNFs for neutral scalar perturbations for the model of
Ref.~\cite{ayonbeato}. The $q=0$ values reproduce those of
Ref.~\cite{iyer2} for the Schwarzschild case.}
\end{table}

\begin{table}[b]
\begin{tabular}{llccc}
 \hline
$l$ ~~~ ~~~& $n$ ~~~~~~& $b=0.1$~~~& $b=0.5$~~~& $b=1$\\
 \hline
0  & 0 & $0.1045-\imath~0.1152$  ~~~& $0.1032-\imath~0.1166$ ~~~& $0.0989-\imath~0.1210$\\
  & 1 & $0.0891-\imath~0.3551$  ~~~& $0.0872-\imath~0.3590$ ~~~& $0.0811-\imath~0.3716$\\
\hline
1   & 0 & $0.2910-\imath~0.0980$ ~~~& $0.2885-\imath~0.0986$ ~~~& $0.2813-\imath~0.1006$\\
   & 1 & $0.2620-\imath~0.3075$ ~~~& $0.2587-\imath~0.3097$ ~~~& $0.2489-\imath~0.3167$\\
   & 2 & $0.2233-\imath~0.5269$ ~~~& $0.2186-\imath~0.5309$ ~~~& $0.2047-\imath~0.5431$\\
   & 3 & $0.1734-\imath~0.7488$ ~~~& $0.1667-\imath~0.7546$ ~~~& $0.1466-\imath~0.7724$\\
\hline
2   & 0 & $0.6749-\imath~0.0965$ ~~~& $0.4791-\imath~0.0973$ ~~~& $0.4677-\imath~0.0990$\\
   & 1 & $0.6601-\imath~0.2924$ ~~~& $0.4584-\imath~0.2976$ ~~~& $0.4451-\imath~0.3031$\\
   & 2 & $0.6345-\imath~0.4942$ ~~~& $0.4259-\imath~0.5068$ ~~~& $0.4095-\imath~0.5166$\\
   & 3 & $0.6018-\imath~0.7012$ ~~~& $0.3854-\imath~0.7208$ ~~~& $0.3648-\imath~0.7351$ \\
 \hline
\end{tabular}
\caption{QNFs for neutral scalar perturbations for the solution of
Ref.~\cite{fabris} obtained for a system of gravity coupled to a
phantom scalar field. The parameters have been fixed in order to
normalize to unity the black hole mass, $c=-3\pi/(2b)$ and $\rho_0=3$
and$b$ has been set to $b=0.1,~0.5,1$.}
\end{table}

\centerline{}
\newpage
\centerline{}
\newpage
\section{Tables of QNFs for charged scalar perturbations}

\begin{table}[ht]
\begin{tabular}{llccc}
 \hline
$l$ ~~~ ~~~& $n$ ~~~~~~& $q=0.1$~~~& $q=0.2$~~~& $q=0.3$\\
 \hline
0  & 0 & $0.1451-\imath~0.1198$ ~~~& $0.1895-\imath~0.1224$ ~~~& $0.2393-\imath~0.1219$\\
  & 1 & $0.0730-\imath~0.3454$ ~~~& $0.0479-\imath~0.3027$ ~~~& $0.0683-\imath~0.2577$\\
\hline
1   & 0 & $0.3267-\imath~0.1011$ ~~~& $0.3655-\imath~0.1036$ ~~~& $0.4082-\imath~0.1051$\\
   & 1 & $0.3005-\imath~0.3144$ ~~~& $0.3369-\imath~0.3189$ ~~~& $0.3729-\imath~0.3200$\\
   & 2 & $0.2540-\imath~0.5292$ ~~~& $0.2517-\imath~0.5192$ ~~~& $0.2479-\imath~0.4841$\\
   & 3 & $0.1635-\imath~0.7248$ ~~~& $0.0923-\imath~0.6242$ ~~~& $0.0938-\imath~0.5115$\\
\hline
2   & 0 & $0.5182-\imath~0.0987$ ~~~& $0.5564-\imath~0.1002$ ~~~& $0.5987-\imath~0.1012$\\
   & 1 & $0.5000-\imath~0.3012$ ~~~& $0.5397-\imath~0.3051$ ~~~& $0.5831-\imath~0.3071$\\
   & 2 & $0.4708-\imath~0.5106$ ~~~& $0.5096-\imath~0.5151$ ~~~& $0.5488-\imath~0.5167$\\
   & 3 & $0.4323-\imath~0.7219$ ~~~& $0.4607-\imath~0.7238$~~~& $0.4791-\imath~0.7197$ \\
 \hline
\end{tabular}
\caption{QNFs for charged scalar perturbations and for the solution of
Ref.~\cite{bardeen,AyonBeato:2000zs}. We normalized $m$ and $e$ to unity.}
\end{table}
\begin{table}[ht]
\begin{tabular}{llccc}
 \hline
$l$ ~~~ ~~~& $n$ ~~~~~~& $q=0.1$~~~& $q=0.2$~~~& $q=0.3$\\
 \hline
0  & 0 & $0.1447-\imath~0.1201$ ~~~& $0.1895-\imath~0.1247$ ~~~& $0.2383-\imath~0.1251$\\
  & 1 & $0.0724-\imath~0.3463$ ~~~& $0.1684-\imath~0.3435$ ~~~& $0.1368-\imath~0.2724$\\
\hline
1   & 0 & $0.3266-\imath~0.1013$ ~~~& $0.3647-\imath~0.1044$ ~~~& $0.4060-\imath~0.1072$\\
   & 1 & $0.3002-\imath~0.3150$ ~~~& $0.3355-\imath~0.3216$ ~~~& $0.3690-\imath~0.3266$\\
   & 2 & $0.2535-\imath~0.5303$ ~~~& $0.2500-\imath~0.5222$ ~~~& $0.2407-\imath~0.4907$\\
   & 3 & $0.1626-\imath~0.7262$ ~~~& $0.0926-\imath~0.6279$ ~~~& $0.0871-\imath~0.5162$\\
\hline
2   & 0 & $0.5181-\imath~0.0989$ ~~~& $0.5559-\imath~0.1010$ ~~~& $0.5972-\imath~0.1031$\\
   & 1 & $0.4998-\imath~0.3017$ ~~~& $0.5388-\imath~0.3075$ ~~~& $0.5807-\imath~0.3132$\\
   & 2 & $0.4705-\imath~0.5116$ ~~~& $0.5081-\imath~0.5193$ ~~~& $0.5448-\imath~0.5268$\\
   & 3 & $0.4317-\imath~0.7234$ ~~~& $0.4581-\imath~0.7295$~~~& $0.4721-\imath~0.7322$ \\
 \hline
\end{tabular}
\caption{QNFs for charged scalar perturbations and for the solution of
Ref.~\cite{dymn2}. We normalized $m$ and $e$ to unity.}
\end{table}

\begin{table}[ht]
\begin{tabular}{llccc}
 \hline
$l$ ~~~ ~~~& $n$ ~~~~~~& $q=0.1$~~~& $q=0.2$~~~& $q=0.3$\\
 \hline
0  & 0 & $0.1459-\imath~0.1199$ ~~~& $0.1929-\imath~0.1224$ ~~~& $0.2473-\imath~0.1215$ \\
  & 1 & $0.0874-\imath~0.3515$ ~~~& $0.0526-\imath~0.3059$ ~~~& $0.0703-\imath~0.2557$\\
\hline
1   & 0 & $0.3275-\imath~0.1012$ ~~~& $0.3689-\imath~0.1038$ ~~~& $0.4175-\imath~0.1054$\\
   & 1 & $0.3014-\imath~0.3145$ ~~~& $0.3407-\imath~0.3194$ ~~~& $0.3828-\imath~0.3227$\\
   & 2 & $0.2559-\imath~0.5292$ ~~~& $0.2573-\imath~0.5199$ ~~~& $0.2462-\imath~0.4930$\\
   & 3 & $0.1704-\imath~0.7238$ ~~~& $0.1008-\imath~0.6282$ ~~~& $0.0830-\imath~0.5105$\\
\hline
2   & 0 & $0.5192-\imath~0.0988$ ~~~& $0.5609-\imath~0.1004$ ~~~& $0.6103-\imath~0.1015$\\
   & 1 & $0.5010-\imath~0.3013$ ~~~& $0.5443-\imath~0.3056$ ~~~& $0.5951-\imath~0.3080$ \\
   & 2 & $0.4719-\imath~0.5108$ ~~~& $0.5145-\imath~0.5160$ ~~~& $0.5620-\imath~0.5184$\\
   & 3 & $0.4335-\imath~0.7222$ ~~~& $0.4660-\imath~0.7249$ ~~~& $0.4943-\imath~0.7257$ \\
 \hline
\end{tabular}
\caption{QNFs for charged scalar perturbations and for the solution of
Ref.~\cite{ayonbeato}. We normalized $m$ and $e$ to unity.}
\end{table}

\end{widetext}

\centerline{}
\vfill\eject
\newpage
\centerline{}
\newpage
\centerline{}
\vfill\eject

\end{document}